\documentclass[11pt,twoside]{article}
\usepackage{asp2010}
\usepackage{epstopdf}
\usepackage[margin=10pt,font=small,labelfont=bf, labelformat=simple]{caption}

\begin{document}

\title{Galaxy pairs as a probe for mergers at $z \sim 2$}
\author{Allison W. S. Man$^1$, Andrew Zirm$^1$, Sune Toft$^1$
\affil{$^1$Dark Cosmology Centre, University of Copenhagen, Denmark}}

\begin{abstract}
In this work I investigate the redshift evolution of pair fraction of a sample of 196 massive galaxies from $z$ = 0 to 3, selected from the COSMOS field.
We find that on average a massive galaxy undergoes $\sim 1.1 \pm 0.5$ major merger since $z=3$.
I will review the current limitations of using the pair fraction as a probe for quantifying the impact of mergers on galaxy evolution.
This work is based on the paper \citet{Man2011}.
\end{abstract}
\vspace{-20pt}

\section{Introduction}
Mergers are thought to be one of the primary drivers of galaxy evolution, at least since $z \sim 3$.
Theories and simulations have shown that mergers are able to transform the structure of a galaxy, and are required for the mass assembly of galaxies.
However, it remains an open issue how important merging is, compared to other physical mechanisms such as cold gas accretion or secular evolution.
Hence there is a motivation to quantify the importance of merger through observations.

Typically merger candidates are identified from photometric catalogues through either pair selection \citep{Bluck2009, Williams2011, Newman2011} or morphology \citep{Lotz2008, Bluck2011}.
Irregular morphologies at high redshift could be due to cosmological surface-brightness dimming or clumpy star formation, and are not always related to merging activity.
Hence we focus on using pair counts as a probe for identifying mergers in this work.
\vspace{-10pt}

\section{Selecting galaxy pairs}
We use a compilation of all public photometric data in the COSMOS field, in 30+ narrow-, medium- and broad-bands covering wavelengths in UV, optical, NIR and mid-IR.
%The parent catalogue is selected in the $i$-band (from Subaru Suprime-Cam), where fluxes are measured within apertures of 3$\arcsec$ in diameter and has a limiting magnitude of $i < 26$.
Photo-$z$'s and stellar population fitting are derived using the EAZY and FAST codes with standard parameters.
% are derived on all entries using the medium- and broad-band catalogue with the code EAZY \citep{Brammer2008}.
%We make use of the BC03 stellar population synthesis model with the FAST code \citep{Kriek2009}, assuming a Chabrier initial mass function (IMF), and fit the SEDs with three different star formation histories: a single stellar population without dust, an exponentially declining model with \textit{e}-folding time of 300 Myr and dust attenuation allowed to be between $A_\textrm{v}$ = 0-4, and a constant star formation model with the same range in attenuation.
%We assume solar metallicity and the Calzetti extinction law.
We constrain ourselves to the area which has $HST$/NICMOS $H_{160}$-band imaging, which covers $\sim 5 \%$ of the COSMOS area.
The observed NIR imaging probes the rest-frame optical wavelength, which is a better tracer of the bulk stellar mass compared to observed optical (rest-frame UV).
There are 196 massive ($M_{\star} \ge 10^{11}~M_{\odot}$) galaxies at $0 \le z \le 3$.

We run \verb"SExtractor" on the NICMOS cutouts of the massive galaxies, and select galaxy pairs with these two criteria:
(1) the massive galaxy has one or more companion within a projected separation of 30 kpc; and
(2) the $H_{160}$-flux ratio of the pair is between 1:4 to 1:1.
Imposing these criteria we find 40 massive galaxies in pairs, in the redshift range of $0 \le z \le 3$.
%The number of galaxies having companions of flux ratio above 1:2 / 1:3 / 1:4 is 20 / 32 / 40 respectively.
%Examples of the cutouts are shown in Figure~\ref{fig:pairs_images}.

%%%%%%%%%%%%%%%%%%%%%%%%%%%%%%%%%%
% NICMOS cutouts of pairs
%\begin{figure}[h!]
%	\centering
%	\includegraphics[angle=0,width=0.5\textwidth]{fig/9cutouts}
%	\caption{The NICMOS $H_{160}$ postage stamps of nine examples of the selected galaxy pairs.
%	The top row shows pairs that were source-confused in the original COSMOS catalogue, but are now resolved in our analysis with the NICMOS imaging;
%	the bottom two rows contain pairs that have individual entries in the catalogue.
%	The IDs and photo-$z$'s of the massive galaxies are labelled on the top left and right hand corners of each panel.
%	The angular scale and the 30-kpc search radius are indicated with the vertical bar and white circle, respectively.
%	For illustrative purpose, the colour coding is scaled logarithmically and the images are smoothed by convolving with a Gaussian PSF of FWHM = 2 pixels ($0.202\arcsec$).
%	The angular scale is shown with the 1$\arcsec$ vertical bar.
%	The white circle overlaid on each map indicates the 30-kpc search radius around each massive galaxy at the centre.
%	}
%	\label{fig:pairs_images}
%\end{figure}
%%%%%%%%%%%%%%%%%%%%%%%%%%%%%%%%%%

The pair fraction is corrected for chance projection using two different approaches:
(1) we perform a Monte-Carlo simulation, in which the positions of all the COSMOS sources are redistributed randomly and we replicate our pair selection; or
(2) we impose an additional criterion: if the companion has a separate COSMOS entry with reliable photo-$z$ (\verb"odds" $\ge$ 0.95), the 3$\sigma$ confidence intervals of the photo-$z$'s must overlap.
\vspace{-10pt}

\section{Results}
\begin{figure}[htbp]
%\begin{minipage}[l]{0.5\linewidth}
	\centering
	\includegraphics[angle=0,width=0.7\textwidth]{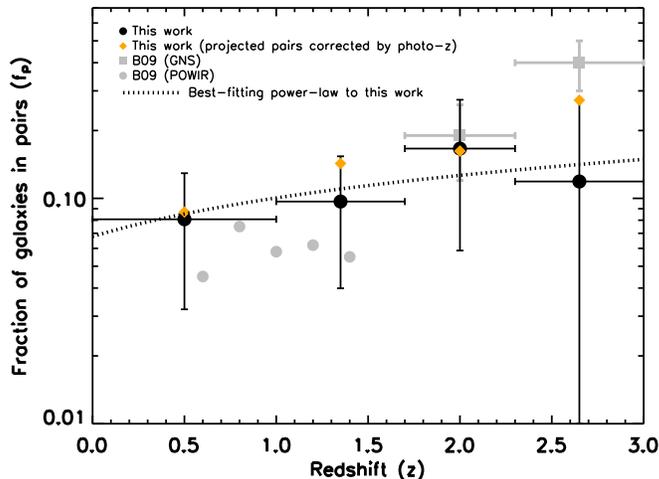}
	\vspace{-15pt}
	\caption{The redshift evolution of the pair fractions $f_\mathrm{p}$, compared to other observations. 
	The black circles denote the $f_{p}$ of our analysis, after statistically correcting for projection contamination.
	The black dotted line show the best-fitting power law to our $f_{p}$, which is of the form $F(z) = (0.07 \pm 0.04) \times (1+z)^{0.6 \pm 0.5}$.
	The orange diamonds denote our $f_{p}$, which we use an alternative approach to correct for projection contamination with the available photo-$z$'s.
	The gray squares and circles represent the $f_{p}$ of \citet{Bluck2009} using the GNS and POWIR data.
	 %Assuming that the uncertainties in the counts follow the Poisson distribution, they are overplotted as the vertical error bars.
	 %The horizontal bars indicate the width of each bin.
	}
	\label{fig:pf_lit}
%\end{minipage}%
\end{figure}

As seen on Figure~\ref{fig:pf_lit},
the pair fraction derived using the photo-$z$ criterion is remarkably consistent with the statistical approach, except at the highest redshift bin in which photo-$z$'s have higher uncertainties and therefore less constraining.
%, but still marginally consistent.
%We fit the observed $f_{p}$ with a power law of the form $F(z) = F(0) (1+z)^{m}$, and find the best fit parameters to be $F(0) = 0.07 \pm 0.04$ and $m = 0.6 \pm 0.5$.

\citet{Bluck2009} and \citet{Williams2011} have performed a similar study with different samples. Our results are consistent with theirs within the uncertainties.
However, readers should be cautioned that \citet{Bluck2009} claim a strong redshift evolution of the pair fraction, whereas \citet{Williams2011} claim a mild (and even diminishing one).
\citet{Lotz2011} demonstrate with simulations that much of the discrepancy can be explained by the different criteria used in the analyses (mass- or luminosity-selected parent sample, projected separation, etc.).
%%%%%%%%%%%%%%%%%%%%%%%%%%%%%%%%%%
% Redshift evolution of pair fraction (vs literature)
%%%%%%%%%%%%%%%%%%%%%%%%%%%%%%%%%%
%%%%%%%%%%%%%%%%%%%%%%%%%%%%%%%%%%
% Number density evolution
%\begin{minipage}[r]{0.5\linewidth}
\begin{figure}[htbp]
	\centering
	\includegraphics[angle=0,width=0.7\textwidth]{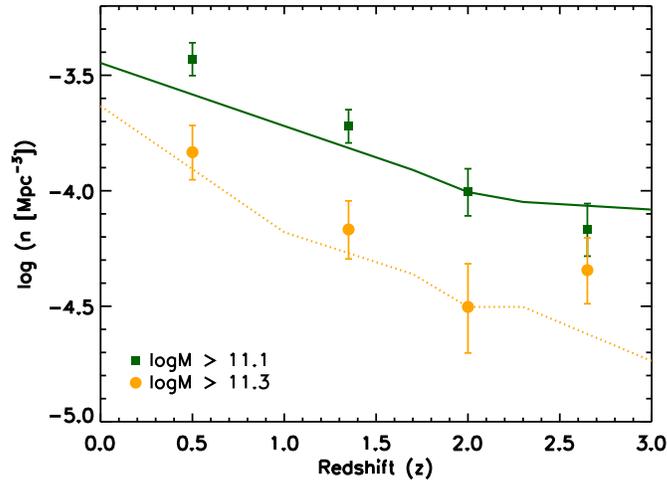}
	 \vspace{-15pt}
	\caption{The redshift evolution of number density of massive galaxies.
	The filled symbols are the observed co-moving number density of massive galaxies from our sample, with mass limits shown in the legend.
	The lines represent the predicted number growth using the observed number density of close pairs, after correcting for projected pairs using photo-$z$.
	The lines are normalized to the observed number density at $z = 2$.}
	\label{fig:numdens}
%\end{minipage}
\end{figure}%

%%%%%%%%%%%%%%%%%%%%%%%%%%%%%%%%%%
Assuming a merging timescale of $\tau = 0.4 \pm 0.2$ Gyr, 
we integrate the pair fraction over our redshift range and the co-moving volume, 
and find that \textbf{a massive galaxy experiences $N_{m} = 1.1 \pm 0.5$ major mergers on average from $z=3$ to 0.}
Observations have shown that massive galaxies grow 3-6 times in size during this epoch,
and it has been estimated that 2-3 major mergers \citep{Bezanson2009} are required to explain this size growth.
Hence we can conclude that \textbf{mechanisms other than major merging are required to explain the size evolution}.
This is not surprising as minor mergers are more frequent and more efficient in growing galaxies' sizes \citep{Naab2009}.

We estimate the number of newly created massive galaxies as follows:
for each pair, we calculate the remnant mass as the sum of the SED masses of the galaxies in the pair.
In the rare case (4 pairs) where the SED mass is not available for the companion galaxy because there is no corresponding entry in the catalogue, 
we use the flux ratio and the SED mass of the primary massive galaxy to estimate the remnant mass.
The number of newly created massive galaxy (N$_\mathrm{created}$) in each redshift bin is calculated by counting the galaxies that cross the mass limit after merging.
The merger-induced increment in the co-moving number density ($\Delta$, in units of Mpc$^{-3}$) is given by: $ \Delta = \frac{\mathrm{N_{created} } \times t_\mathrm{elapsed}}{ V_\mathrm{co-moving} \times  \tau} $
where $t_\mathrm{elapsed}$ is the time elapsed within the redshift bin.
Normalizing the number density of massive galaxies to the observation at $z=2$,
the results are compared with the observed number density of massive galaxies above these mass limits, 
as shown in Figure~\ref{fig:numdens}.
Considering the $\sim$ 0.2 dex uncertainty in the number density growth due to counting statistics,
the slope of the number growth is remarkably consistent with the observed number density.
The agreement between our estimated merger-induced number density growth and the observed number density supports the idea that \textbf{major mergers are sufficient to explain the number density evolution of massive galaxies from $z \sim 2.3$ to 0}.
\vspace{-10pt}

\section{Discussion}
In the near future, I will make use of the deep, large-area UltraVISTA survey in the COSMOS field to analyze the pair fraction.
In combination with the Ultra-Deep Survey (UDS), we should be able to trace the major and minor merger fraction at $2 < z < 3$ to less than 20\% uncertainty.
We will be able to investigate the effect of environment:
one may expect higher pair fractions in overdense regions,
but the relative velocities between the galaxy pairs might also be higher,
such that the probability that the pair is undergoing a merger is low compared to the field.

However, even if we had perfect measurements of the merger fraction,
we are still unable to determine the observability timescale of the merger to less than 50\% uncertainty.
There are attempts to solve it through merger simulations, but the problem itself is complex:
the timescale is sensitive to the initial conditions of the merger \citep{Hopkins2010}, such as mass ratio, gas fraction, dust content, relative velocity, orbital parameters, viewing angle, etc.
Some of these parameters may evolve with redshift and environment.
Hence more work needs to be done from the theory and the simulation fronts, to interpret statistically how the observed frequency of mergers impact the evolution of galaxies.
\vspace{-10pt}

\section{Conclusions}
\begin{itemize}
	\item The pair fraction evolves only mildly with redshift since $z=3$.
	\vspace{-10pt}
	\item A massive galaxy experiences $N_{m} = 1.1 \pm 0.5$ major mergers on average since $z=3$. 
	This amount of major merging is sufficient to explain the number of newly formed massive galaxies, but insufficient to explain the observed size evolution.
	This hints that massive galaxies undergo other mechanisms that is efficient in puffing up their sizes, but not so important in growing the mass.
	Minor merging is compatible with this scenario.
	\vspace{-10pt}
	\item It remains a challenge to constrain the observability timescale of mergers due to the complexity of the problem.
	Its importance oughts to be highlighted, as it is a crucial quantity for interpreting the impact of mergers on the growth of galaxy population as a whole.
\end{itemize}
\vspace{-10pt}

\acknowledgements
To everyone who organized or participated in the conference on Galaxy Mergers in an Evolving Universe, thank you for the great time!
I also thank my co-authors, Stijn Wuyts and Arjen van der Wel, and the support from Dark Cosmology Centre, for the completion of this project.
\vspace{-10pt}

\bibliographystyle{asp2010} % style aa.bst

\end{document}